\documentclass[]{interact}

\usepackage{epstopdf}
\usepackage{apacite}

\usepackage{siunitx}
\usepackage{subcaption}
\usepackage{color}
\usepackage{tabularray}
\usepackage{algorithm}
\usepackage{algorithmic}
\UseTblrLibrary{diagbox}
\usepackage{graphicx}
\usepackage{booktabs}
\usepackage{multirow}
\usepackage{array}
\usepackage{longtable}

\usepackage[longnamesfirst,sort]{natbib}

\bibpunct[, ]{(}{)}{;}{a}{,}{,}
\renewcommand\bibfont{\fontsize{10}{12}\selectfont}

\theoremstyle{plain}

\theoremstyle{definition}

\theoremstyle{remark}

\begin{document}


\title{Interactive Content Diversity and User Exploration in Online Movie Recommenders: A Field Experiment}

\author{
    \name{
        Ruixuan Sun\textsuperscript{a}\thanks{CONTACT Ruixuan Sun Email: sun00587@umn.edu}
        Avinash Akella \textsuperscript{a}
        Ruoyan Kong \textsuperscript{a}
        Moyan Zhou \textsuperscript{a}
        and Joseph A. Konstan \textsuperscript{a}
    }
    \affil{\textsuperscript{a}Grouplens Research, University of Minnesota; 5-244 Keller Hall, 200 Union Street SE; Minneapolis, Minensota, United States}
}

\maketitle

\begin{abstract}

Recommender systems often struggle to strike a balance between matching users' tastes and providing unexpected recommendations. When recommendations are too narrow and fail to cover the full range of users' preferences, the system is perceived as useless. Conversely, when the system suggests too many items that users don't like, it is considered impersonal or ineffective. To better understand user sentiment about the breadth of recommendations given by a movie recommender, we conducted interviews and surveys and found out that many users considered narrow recommendations to be useful, while a smaller number explicitly wanted greater breadth. Additionally, we designed and ran an online field experiment with a larger user group, evaluating two new interfaces designed to provide users with greater access to broader recommendations. We looked at user preferences and behavior for two groups of users: those with higher initial movie diversity and those with lower diversity. Among our findings, we discovered that different level of exploration control and users' subjective preferences on interfaces are more predictive of their satisfaction with the recommender.

\end{abstract}


\section{Introduction}

To deliver personalized recommendations, it is essential to understand end users' information-seeking goals \citep{mcnee2006making}. The recommendation strategies should adapt accordingly based on the scenario. For instance, when searching for a local football game's score on Google, users want a narrow list of results to quickly find the needed information. In contrast, if a user wants to buy a gift for children but has no specific idea in mind, the recommender system should support the ability to explore a broad range of available choices. Therefore, a mapping is required between recommendation strategies and information-seeking scenarios.

Even without recommendations, people tend to consume narrower content over time \citep{nguyen2014exploring}. One extreme phenomenon that can happen under such circumstances is pigeonholing \citep{mcnee2006making}, where users feel that the recommendations are too narrow and too similar to what they have already seen. For example, if a user has only shown interest in home improvement videos, a recommendation model might exploit that preference and display mostly DIY project videos to this user. This concept goes back to a long way. In the earlier years, we see \citeauthor{ziegler2005improving} suggested allowing users to adjust the level of recommendations, and \citeauthor{mcnee2006making} defined pigeonholing as an important personality in Human-Recommeder Interaction (HRI) analytical model. Later, \citeauthor{ekstrand2015letting} have run field experiment to let users choose their recommender algorithms and analyzed the relationship between recommendation diversity and user preferences of certain algorithms. In industry, practitioners and journalists have also identified the harmful impact of the pigeonholing phenomena. Youtube is critized for its polarization towards ideologically narrower content recommendation over time\citep{brown2022echo, lutz2021examining}, and Google News has bias in over-representing certain news outlets and brings concern of viewpoint diversity \citep{haim2018burst}.

On the other end of the spectrum, we have encouraging user exploration, which some studies have shown that broadening user exploration on less exposed content with reinforcement learning models can improve the long-term user experience \citep{chen2021values}. Some previous works have also considered both exploit and explore strategies, demonstrating the effectiveness of a multi-bandit strategy in improving user engagement \citep{mcinerney2018explore}. Although many studies have investigated model-based user exploration and proposed algorithms to provide a certain level of novelty and serendipity \citep{chen2021values, zhang2012auralist, kaminskas2016diversity}, few have considered recommendation usefulness and how it depends on users and their specific circumstances.

In this study, we experiment with providing users with more recommendation possibilities via interactive recommendation interfaces, prioritizing the user exploration experience. We found that access to broad recommendations not only improved user engagement and general satisfaction but also provided users with opportunities to step out from their regular consumption zones and achieve different exploration goals. In the following section, we discuss related works, followed by a formative interview study, an online field experiment, and a post-survey. Results and discussion will be shared afterward. Finally, we wrap up this paper with a conclusion and future implications.

\section{Related Work}

\subsection{User Exploration, Engagement, and Satisfaction}
One of the important dynamics of Human-Recommender Interaction (HRI) is how users explore items and consequently, engage with the system. Previous research has identified that traditional recommendation algorithms such as collaborative filtering (CF) or content-based filtering (CBF) incline to pigeonhole users into items similar to the ones they have already expressed interest in \citep{nguyen2014exploring}. That mechanism potentially deprives users of being exposed to more diverse content that could broaden their consumption, and quite a few papers have discussed the bias \citep{baeza2020bias} or mitigation methods \citep{gao2022mitigating} to deal with it. Many previous studies have analyzed the strong correlation between user exploration and satisfaction levels. For example, \citeauthor{schnabel2018short} found that short-term user satisfaction varied depending on different levels of exploration~\citep{schnabel2018short}. \citeauthor{chen2021values} demonstrated how the explore-exploit mechanism of Reinforcement Learning (RL) bandit models helps improve long-term user experience via recommendation quality measurements~\citep{chen2021values}. Past findings have also utilized surveys and interviews to understand user exploration. \citeauthor{liang2019recommender} pointed out that user-centric self-reported behavior was an important measurement for preference and goal changes during their interaction with recommenders~\citep{liang2019recommender}. \citeauthor{garcia2018understanding} found out that user behavior change and interaction on individual tracks can inform how satisfied they are with music discovery~\citep{garcia2018understanding}.

Encouraging users to stay longer and interact more actively with the recommender system is another challenge faced by researchers and practitioners. Previous works have focused on user engagement prediction \citep{said2014recommender} and optimizing long-term user engagement with bandit-based approaches \citep{wu2017returning,zou2019reinforcement}. \citeauthor{zhao2018explicit} pointed out that engagement is a binary notation and usually positive and negative engagement are associated together towards user general experience \citep{zhao2018explicit}. Recently, \citeauthor{kalimeris2021preference} discovered that user engagement can be increased with mitigated echo chamber effect and increased diversity on content\citep{kalimeris2021preference}. In general, researchers agree that user engagement drives long-term user satisfaction with recommender platforms \citep{mladenov2020optimizing,zhao2018explicit,zou2019reinforcement}.

\subsection{Recommendation Diversity}
As many previous findings indicated~\citep{chen2021values,kaminskas2016diversity,zhao2016group,hurley2011novelty,kong2022multi}, accuracy is not the only criterion for judging the success of a recommender system. Other factors such as recommendation novelty \citep{castells2021novelty}, serendipity \citep{chen2019serendipity}, or transparency \citep{shin2019role} also highly impact user engagement and general satisfaction of the system. However, compared to values like serendipity, which requires a good estimate about if users know about and like the recommended item, or transparency, which has no one universal definition and can be perceived differently by individual users, recommendation diversity is more objective and have several commonly recognized formula for its definition \citep{ekstrand2014user, vargas2014coverage, yuan2020attention}.

Content diversity of recommended items is usually a helpful metric to evaluate recommendation quality. In earlier years, \citeauthor{ziegler2005improving} proposed intra-list similarity (ILS) to measure topic diversification in recommendation lists and demonstrated its positive impact on user satisfaction~\citep{ziegler2005improving}. Later, \citeauthor{nguyen2014exploring} revealed how CF-based recommender systems can expose users to narrower content over time, and how user consumption can impact future diversity and rating tendency~\citep{nguyen2014exploring}, or vice versa~\citep{aridor2022economics}. \citeauthor{kaminskas2016diversity} evaluated multiple diversity measurements and summarized general techniques for increasing recommendation diversity~\citep{kaminskas2016diversity}.

However, user perception is a rather subjective notion, therefore not usually aligned with what the measurements suggest. \citeauthor{ferwerda2016influence} looked into how general personality traits relate to preferences of different levels of diversity \citep{ferwerda2016influence}. \citeauthor{tang2022preference} found out the historical diversity traits of users can significantly impact how they accept new diverse recommendations \citep{tang2022preference}. Relevant findings were also presented by \citeauthor{karumur2018personality}, who discovered that users' personality traits have also played a significant role in their engagement intensity and activity types \citep{karumur2018personality}. The implementation of the diversity metric also matters. \citeauthor{jesse2022intra} tested different ways of implementing ILS and indicated that individual validation of the diversity metric choice is necessary for a given application \citep{jesse2022intra}. \citeauthor{ge2013bringing} brought up another point about the ordering of diverse items in the top-n list, which can significantly affect user-perceived diversity \citep{ge2013bringing}.

\subsection{Motivation and Gap}
Many pieces of literature have suggested that accommodating various user exploration needs and promoting positive engagement with recommenders is valuable, and adjusting the diversity of content is one of the most commonly used approaches. However, we have identified a gap in our understanding of how and why diversity changes work, or do not work, for a specific user population on a given platform. To deepen our understanding, we propose a mixed-method study that begins with a semi-structured formative interview to understand real users' current practices and needs. This is followed by a new diversity-centric exploration interface design and a factorial online field experiment with new treatments. Finally, we complement the log interaction data with subjective user feedback through post-surveys to generate a thorough analysis. We believe that this comprehensive end-to-end study, beginning with understanding users and ending with evaluating the same group of users with what they desire, can unveil a realistic relationship between content diversity and user exploration behavior. In a further step, we also hope to bring new perspectives on how user engagement and general satisfaction can be impacted by the new exploration treatment we introduce.

\section{Formative Interview}

To better understand users' information-seeking experience and their preferences for new interactive recommendation interfaces, we conducted semi-structured online interviews \citep{barriball1994collecting} with 17 active\footnote{Active users were considered as those who have logged in to the MovieLens website over 12 times and rated over 20 movies in the year of 2021} users from an online movie recommender system called MovieLens (https://movielens.org/)\footnote{MovieLens is a non-commercial, personalized recommender that gathers users’ ratings on movies they have watched and provides predictions of new movies users might like to watch with different recommendation algorithms. The authors would like to thank the MovieLens team for their support and resources}. We recruited volunteer users through emails. In the recruitment message, we briefly summarized the interview purpose and asked them to sign up for a 30-minute Zoom interview with one researcher. The interview and the following experimental procedure were determined to be a non-human subject study by our local institutional review board. Inspired by findings from previous works~\citep{garcia2018understanding,nguyen2014exploring,taijala2018movieexplorer}, we summarize the purpose of the interview with two RQs:

\textbf{\emph{RQ1}}: What are users' goals and attitudes towards recommendation exploration and pigeonholing?

\textbf{\emph{RQ2}}: What are users' preferences for recommendation interfaces in terms of exploration?

\subsection{Interview Procedure}

In total, we asked 7 questions related to our research questions. Detailed questions are attached in the Appendices. After obtaining users' consent, we collected raw audio recordings, which we analyzed with a data-driven approach guided by the Grounded-Theory Method (GTM) \citep{muller2014curiosity}. First, three researchers involved in this study were informed about the Glaser method \citep{glaser2004remodeling}. Once they had gained sufficient background knowledge, each researcher conducted transcription and open coding separately on their interview scripts. In total, we generated 247 open codes from 17 participants. Then, each of the three researchers read the codes generated by the other two to validate mutual agreement on the code quality. If ambiguity was encountered, all researchers gathered again to discuss it until the disagreement was resolved. In the final step, all researchers met and conducted a thematic analysis~\citep{terry2017thematic} by constructing an affinity map with open codes that had similar meanings. The note clusters were iteratively refined until all researchers reached an agreement on the granularity and accuracy of each theme. Finally, researchers discussed the themes, corresponding novel findings that emerged from each cluster, and summarized four major findings in the section below. We referred to \citeauthor{jin2023collaborative}'s interview steps \citep{jin2023collaborative} in the entire procedure.

\subsection{Interview Findings}

\subsubsection{Usage Frequency and Patterns} Out of the 13 users who could clearly remember their usage frequency, 10 reported that they checked MovieLens for movie recommendations at least once a month. Most users preferred to consume general recommendations from the home page or specific genres rather than searching for specific movies. Our interview also revealed an interesting behavior called "stocking," where some users saved a list of interesting movies during one recommendation session and watched them over the following weeks or months, returning to MovieLens for new batches of recommendations once they had exhausted their current list.

\subsubsection{User goals} Regarding their movie search goals, the majority of users expressed interest in receiving recommendations for movies they already know or ones that are similar to those they have seen before. 10 out of 17 users expressed a desire for MovieLens to suggest different movies that they have not heard of but would enjoy based on their tastes. Additionally, 7 out of 17 users stated that they wanted to find movies that matched specific preferences, such as the genre, director, or if the movie was recently released and still showing in nearby theaters.

\subsubsection{Pain Points} We found two major pain points in user-recommender interactions: pigeonholing and transparency. 

\textbf{\emph{Pigeonholing}} 12 out of 16 users felt they had experienced pigeonholing before. And for those who expressed explicit attitudes toward this phenomenon, 6 out of 10 said pigeonholing was a neutral experience and that either narrow or broad recommendations can be valuable based on the context. One user thought the experience was negative and tried to get rid of it by switching to different recommender algorithms in the settings \footnote{MovieLens allows users who rated over 15 movies to unlock the feature of selecting one of four different algorithms for movie recommendation. More details can be found in section \ref{section 4.2}}. Two users preferred broad recommendations with movies they'd never heard about, and one user was suspicious of how the system could define similar or different movies: \emph{"the notion of similarity was more than just genre or actions."}

\textbf{\emph{Transparency}} 3 users mentioned that they would like to have more transparency in the reasoning behind their recommendations. On the other hand, one user said they felt the recommendation was transparent enough since they could switch between different algorithms \footnote{More details of the existing algorithms on MovieLens can be found in section \ref{section 4.2}.} and test out the difference in generated content. 

\subsubsection{Recommendation Interface} We designed a set of new interface options via text description for users to select from, which either displayed the current exploration information or offered more interactivity to adjust the recommendations for users. More details can be found in the interview script in Appendices. Our primary focus was on enhancing the diversity of content and providing methods to escape from pigeonholing scenarios. Most users indicated that they wanted to explore more options. One of the most popular options was a new carousel that presented broad recommendations with choices that users may not have considered before. This option appealed to almost half of the participants. Another popular choice was to adjust the level of broadness using an interactive slider bar. Users recommended including a precision level that would allow them to switch between different reproducible recommendations. Other popular suggestions included the addition of more filters and basing recommendations on short-term profiles rather than long-term profiles, which was similar to what \citeauthor{taijala2018movieexplorer} did with MovieExplorer \citep{taijala2018movieexplorer}.

\subsection{Preliminary Disccussion}\label{pre_discussion}

In this section, we answer the first two RQs with interview findings and propose the last research question.

\emph{\textbf{RQ1}: What are users' goals and attitudes towards recommendation exploration and pigeonholing?}

The formative interview results revealed clear attitudes and goals for user interaction with the movie recommender. Most users still want to find personalized movies that suit their tastes when receiving recommendations. Additionally, many of them have experienced being pigeonholed, but they did not necessarily consider it a negative experience. Some users might feel that narrow recommendations can bring more personalized and shortlisted candidates that they can efficiently select, while others prefer to get broader choices. Observing this interesting split of two types of users with contrasting diversity tastes, we were inspired to split users into binary groups in a later field experiment based on their historical rating diversity profile. We were curious to analyze how users with opposite diversity preferences react to new recommendation controls differently. 

\emph{\textbf{RQ2}: What are users' preferences for recommendation interfaces in terms of exploration?}

\begin{figure}
    \centering
    \includegraphics[width=0.49\textwidth]{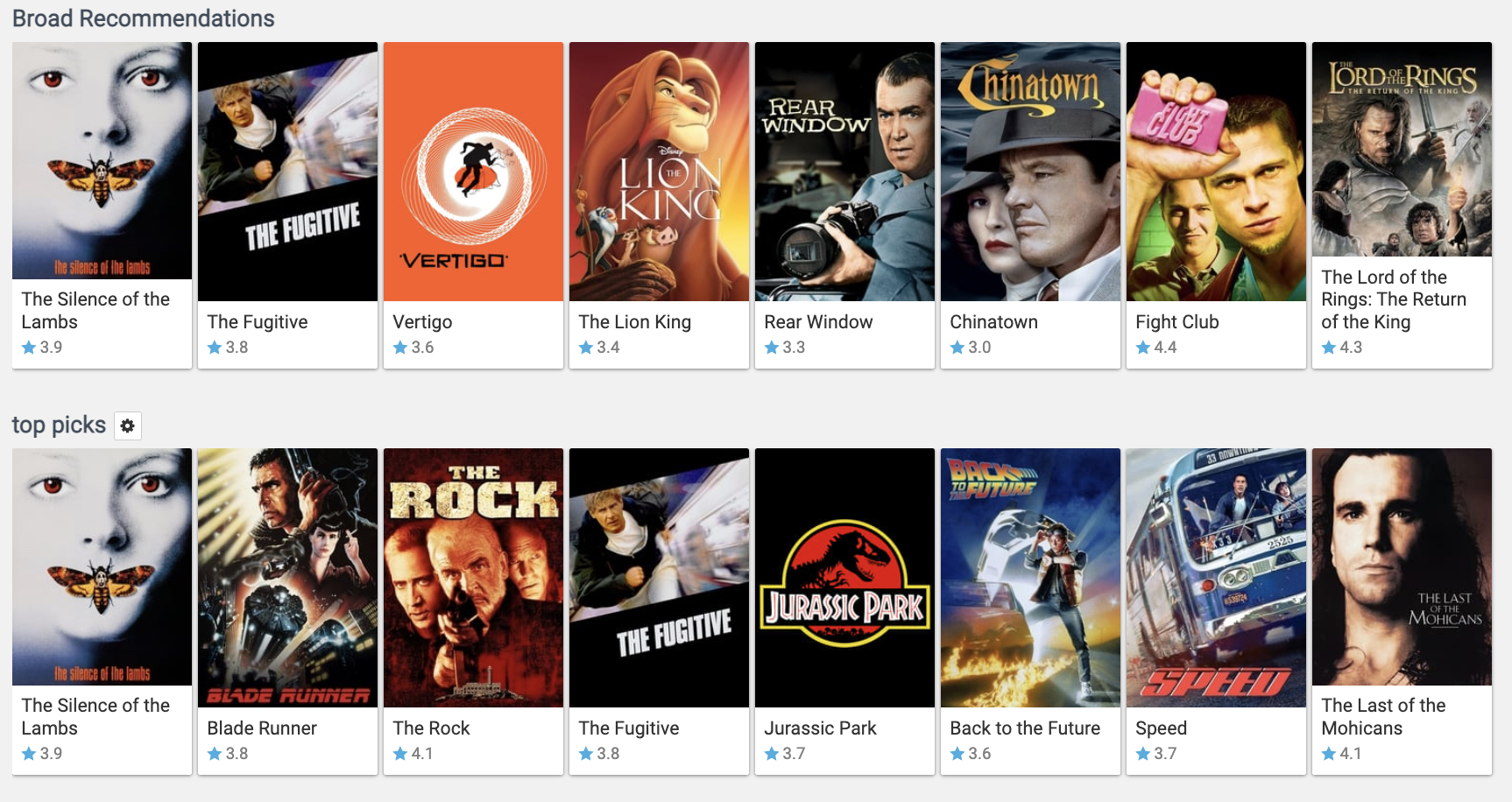}
    \includegraphics[width=0.49\textwidth]{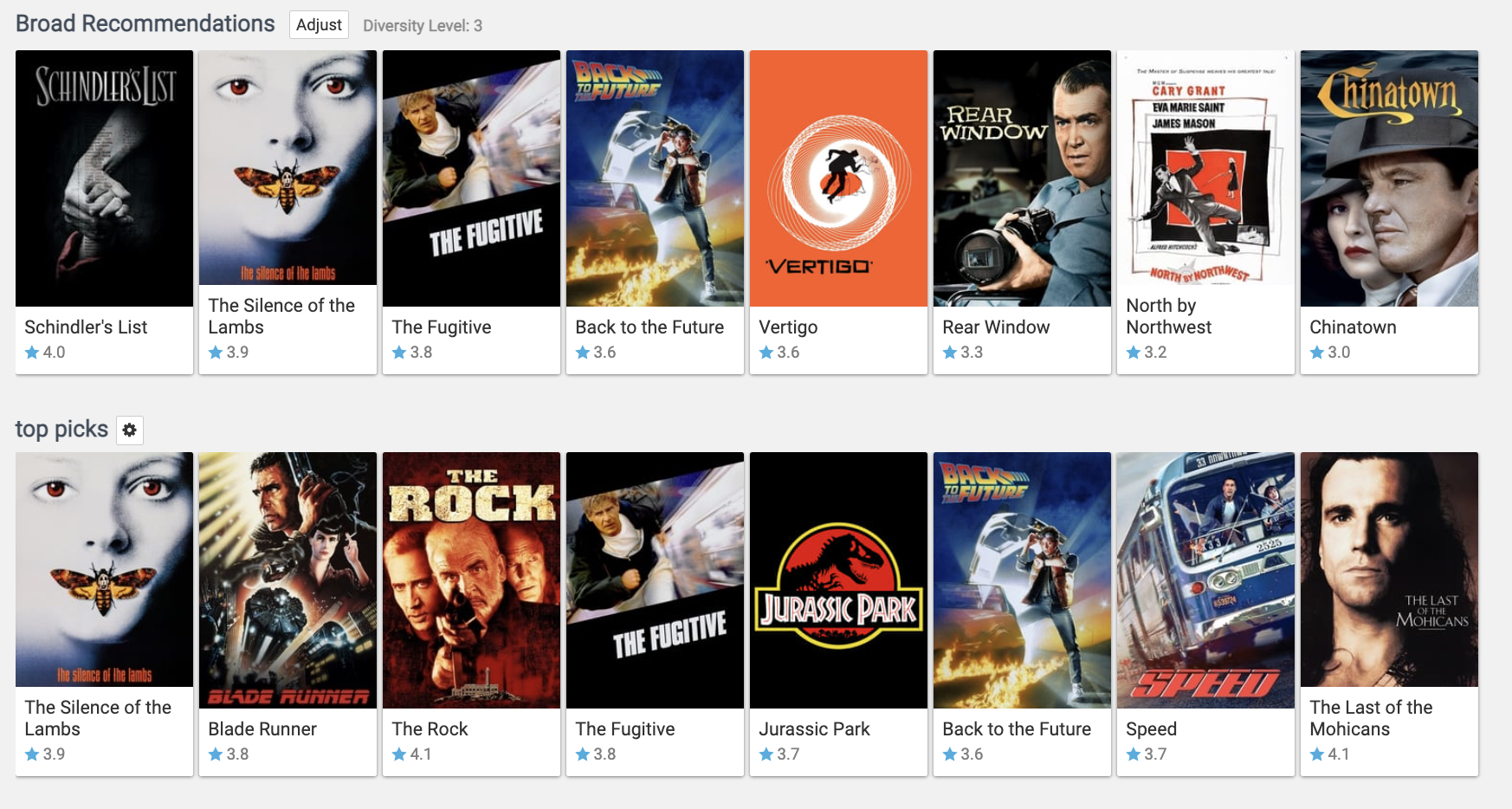}
    \caption{Home page UIs of BRC and BRC+DS. Left: BRC, users can click the "Broad Recommendations" header into the detail recommendation page for more movies; Right: BRC+DS, with an extra "Adjust" button and current diversity level indicator, default to level 3 at the beginning of each new login session. Users can click either the "Broad Recommendations" header or the "Adjust" button into the detail recommendation page.}
    \label{fig1}
    \end{figure}

\begin{figure}
    \centering
    \includegraphics[width=0.49\textwidth]{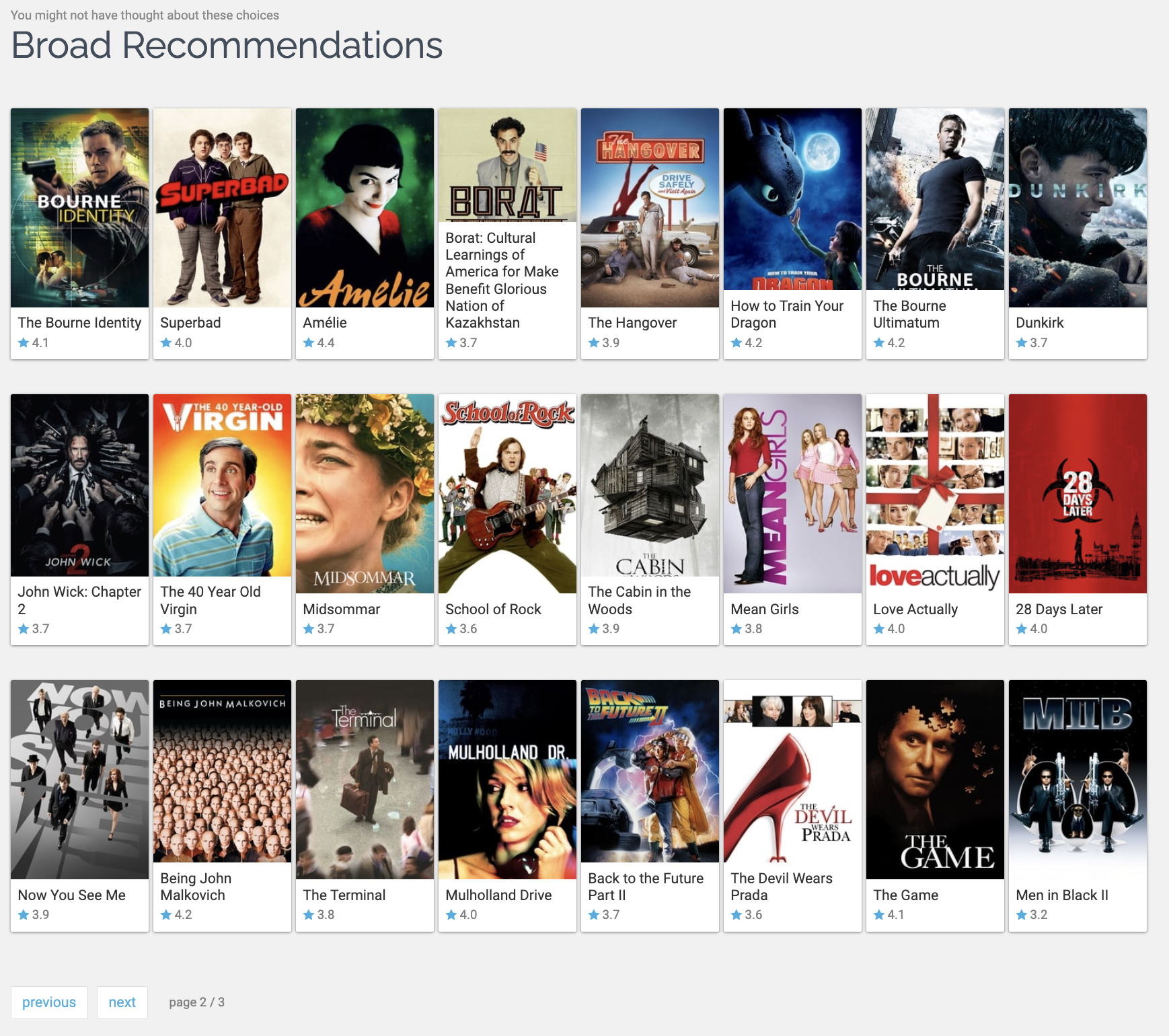}
    \includegraphics[width=0.49\textwidth]{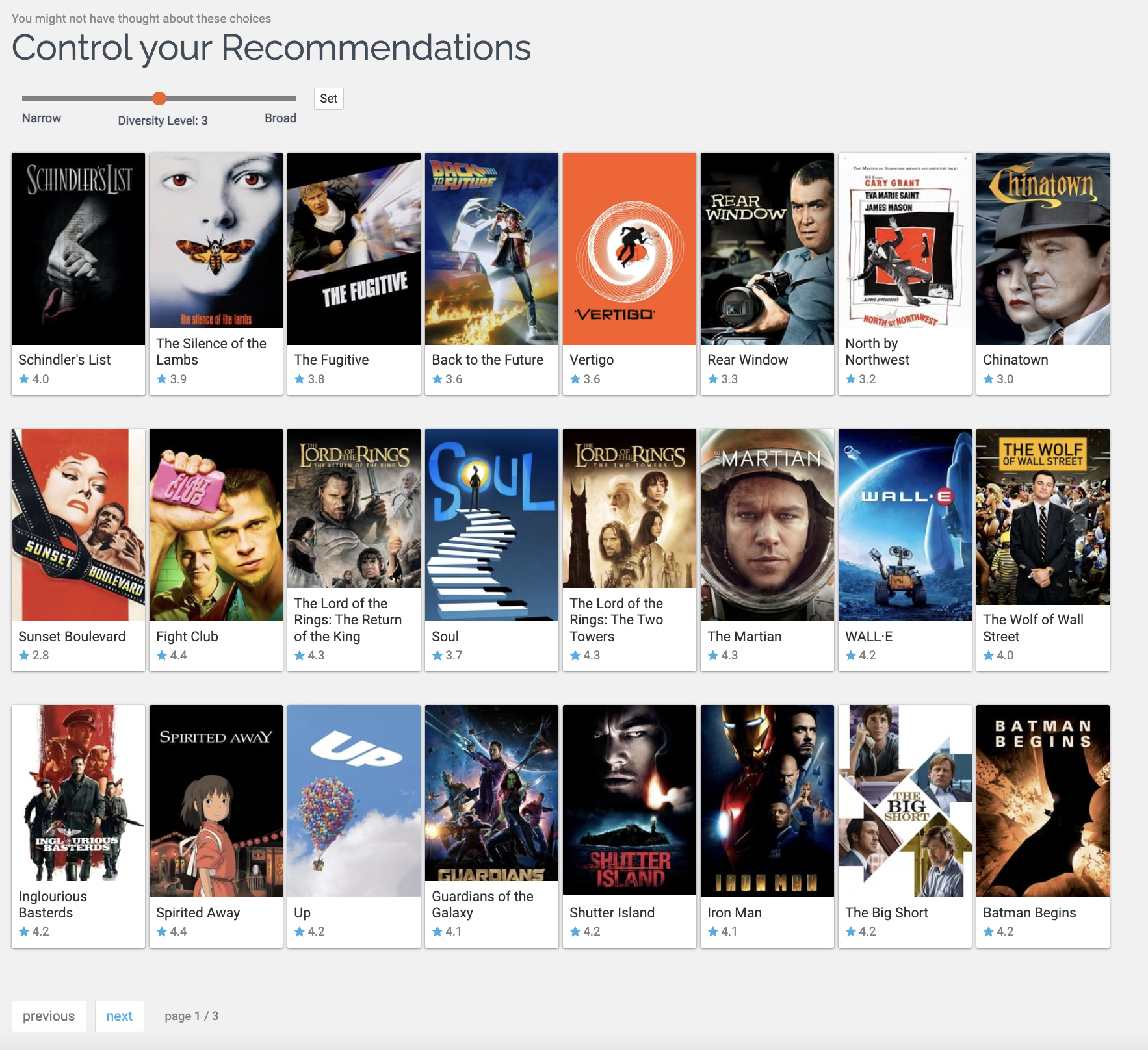}
    \caption{Recommendation detail page UIs of BRC and BRC+DS. Left: BRC that provides the most diverse and personalized set of content corresponding to level 5 on BRC+DS interface; Right: BRC+DS that enabled users to adjust their recommendation content diversity by dragging the slider bar. Users were provided with 5 different levels of diversity to interact with, and the default level was set to 3 when they entered the page during each new login session. When a user selected a diversity level and clicked the "Set" button, the page was refreshed with a new set of movie recommendations based on the indicated diversity level.}
    \label{fig2}
    \end{figure}

Based on user preferences gathered from interviews and technical feasibility assessment, we developed two new exploration interfaces: 1) A broad yet personalized recommendation carousel with more content diversity compared to the "top picks" carousel. We call it the Broad Recommendation Carousel (BRC); 2) A slider that allows users to interact with and select different levels of content broadness on a recommendation page. The entry point to the slider page is also through a BRC. We call it the Broad Recommendation Carousel + Diversity Slider (BRC+DS). Content-wise, movies in BRC could be seen as the "most diverse" (level 5) version of the BRC+DS without interaction features. Both new interfaces are designed as add-ons to the existing UI and are placed at the top level of the home page. Details of example content can be found in Fig. \ref{fig1} and Fig. \ref{fig2}. At each carousel on the home page, users can choose whether to click into the detail page for more specific recommendations or not. Note that we include the term "diversity level" under the slider of BRC+DS interface since we considered the transparency issue some users brought up during the interview. Rather than using a vague name to avoid potential user-perceived bias, we chose to directly tell users what the interface aimed to provide to ensure the best clarity of our interface.

Based on the results obtained from \emph{RQ1} and \emph{RQ2}, we realized that there is a need to establish a stronger connection between users' past consumption patterns and their future exploration behaviors on new interfaces in terms of content diversity flexibility. It is important to note that many previous studies \citep{chen2021values,zhao2018explicit,kalimeris2021preference,zou2019reinforcement} have emphasized the significance of user engagement and satisfaction in evaluating the success of recommender systems. Furthermore, the long-term user experience is considered the ultimate goal in human-centric recommender systems \citep{chen2021values,wu2017returning,zou2019reinforcement,mladenov2020optimizing,zhao2018explicit}. Therefore, we propose the following research question to evaluate the overall merit and benefits of the new recommendation interface in three sub-questions:

\textbf{\emph{NEW RQ3}}: How can the new recommendation interfaces affect users' exploration, engagement, and satisfaction for those with different diversity consumption habits?

\section{Experiment Design}

\begin{figure}[!htp]
    \centering
    \includegraphics[width=0.32\textwidth]{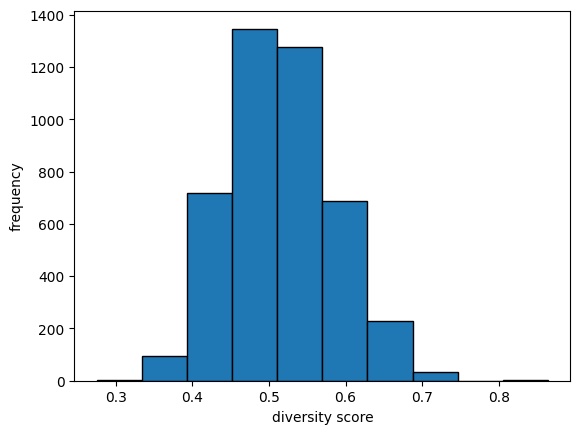}
    \includegraphics[width=0.32\textwidth]{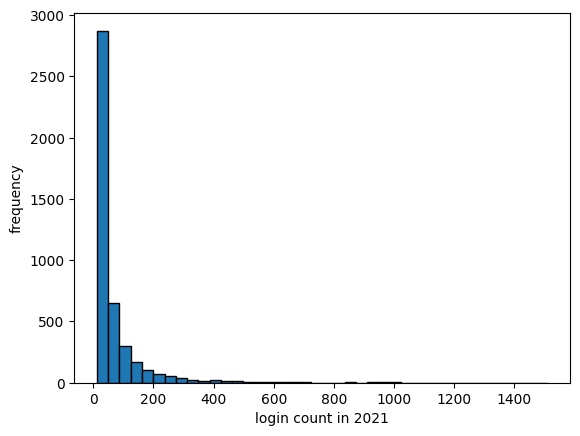}
    \includegraphics[width=0.32\textwidth]{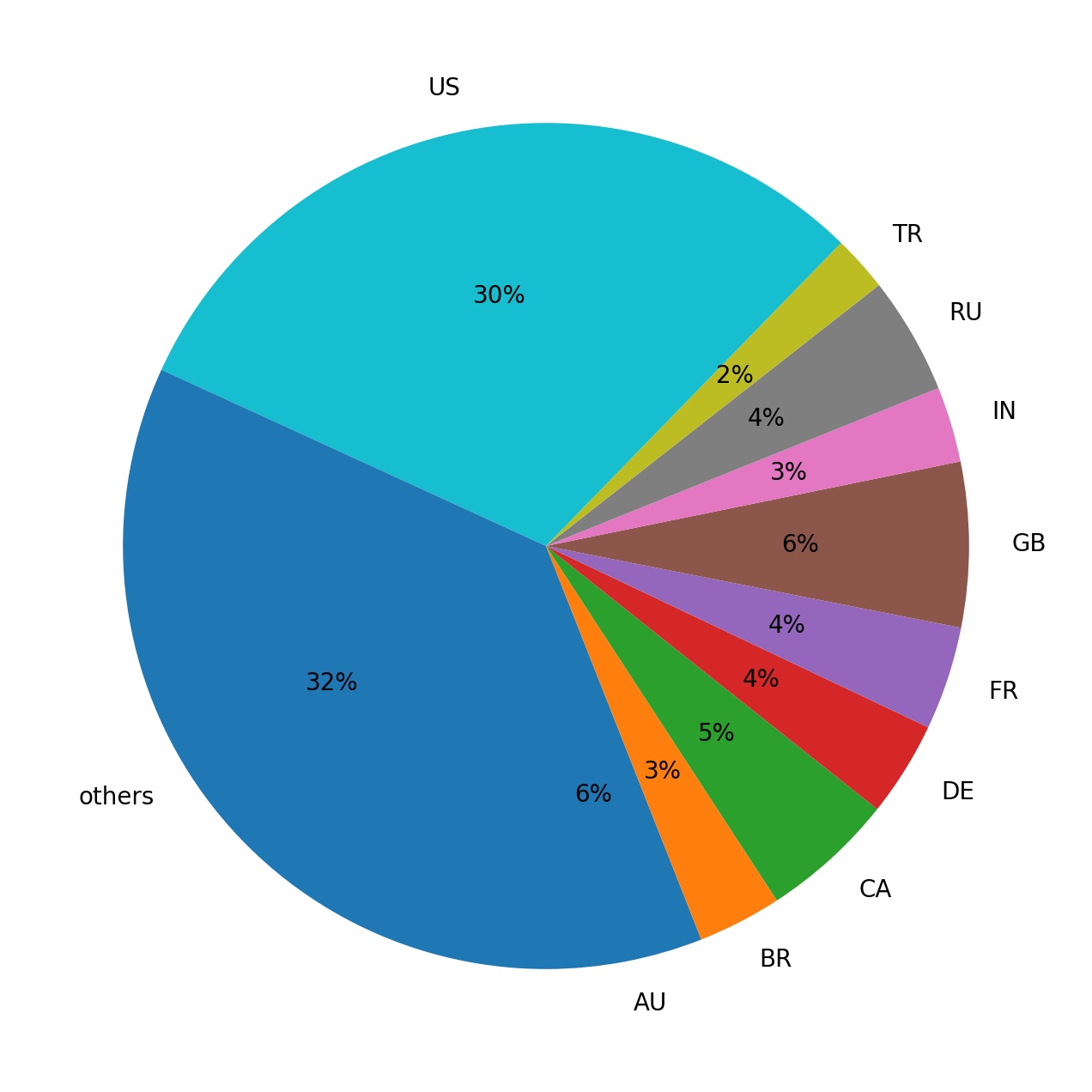}
    \caption{Experiment participant characteristic plots, Left: users' historical rated movie diversity calculated with their tag genome information with our diversity formula \ref{eq:ils_formula}; Middle: users' total number of logins in the year of 2021; Right: users' self-reported country of residence during their registration of account. Note that the residence information collection is not mandatory, therefore we only plot based on available and valid country names.}
    \label{user_characteristics}
    \end{figure}

We designed a 2x3 factorial between-subject experiment with active users in 2021, based on the top new exploration interfaces voted by users. The same selection criteria used for formative interviews were followed, requiring users to have logged in at least 12 times and rated at least 20 items in 2021. Since our system does not collect users' demographic information to protect user privacy, we take an alternative and report participant characteristics based on their consumption diversity, login frequency, and self-reported country of residence in Fig. \ref{user_characteristics}.

\textbf{\emph{Two factors:}} As observed during the formative interviews, users expressed binary preferences regarding exploration goals. To better quantify this pattern, we divided qualified active users into two equal-sized factor groups based on their historical rating diversity: diverse (D) and non-diverse (ND) groups. The diversity threshold for splitting groups was calculated as the median of diversity scores from all users, where each individual score was measured with the average pairwise distance between items' \emph{Tag Genome} feature \citep{vig2012tag} \footnote{More details of tag genomes can be found in section \ref{re-ranking}} in the list of items the user had historically consumed \citep{smyth2001similarity}. We used Pearson correlation \citep{freedman2007statistics} as the distance metric.

    \[
        Diversity(R) = \frac{\sum_{i \in R} \sum_{j \in R\setminus{\{i\}}} dist(i, j)}{|R|(|R|-1)} 
        \label{eq:ils_formula}
        \tag{1}
    \]

\textbf{\emph{Three treatments:}} For each user in both diversity groups, we randomly assigned them to one of three recommendation interfaces, which included one unchanged control interface.

The experiment ran for six weeks, from November 4th, 2022, until December 16th, 2022. During this period, 5007 users visited the website, and 1859 users were invited to join the experiment based on our activity criteria. Overall, 1859 users were assigned to the six (2x3) cohorts, with each cohort having approximately 300 users (detailed numbers are displayed in Table \ref{table1}).

\begin{table}
\centering
\resizebox{\linewidth}{!}{%
\begin{tabular}{cccc} 
\toprule
User Group Treatment                                          & \begin{tabular}[c]{@{}c@{}}Control\\(No Interface Change)\end{tabular} & \begin{tabular}[c]{@{}c@{}}Broad Recommendation \\Carousel (BRC)\end{tabular} & \begin{tabular}[c]{@{}c@{}}Broad Recommendation Carousel\\+ Diversity Slider (BRC + DS)\end{tabular}  \\
\begin{tabular}[c]{@{}c@{}}Diverse \\(n=873)\end{tabular}     & n=281                                                                  & n=288                                                                         & n=304                                                                                                 \\
\begin{tabular}[c]{@{}c@{}}Non-diverse \\(n=986)\end{tabular} & n=328                                                                  & n=339                                                                         & n=319                                                                                                 \\
\bottomrule
\end{tabular}
}
\caption{2x3 Between-Subject Experiment Bucket Split, n indicates actually enrolled users during the experiment.}
\label{table1}
\end{table}

\subsection{Information Window}

Prior to accessing the new recommendation interface, participants were directed to an information window after logging in. We also had a control subgroup for each diverse and non-diverse group as a baseline comparison. In addition to describing the two new types of interfaces, we informed the control groups that the content of the existing top-picks carousel would be different, even if no changes were made. We designed in this way to minimize potential bias for users whose activity level might be impacted by seeing the information window and new interface features. The message in each window was customized for each interface: 

\emph{"Dear MovieLens user, we're experimenting with changes related to the movies we show users, particularly -- }

\indent \textbf{BRC}: \emph{a new carousel that you can find above the top picks carousel on the home page. You may see different content than you're used to.}

\indent \textbf{BRC+DS}: \emph{a new page with a slider bar control that you can enter by clicking the carousel header or "adjust" button next to the top carousel. You will see 5 levels of diversity to toggle with.}

\indent \textbf{Control}: \emph{the top-picks carousel. You may see different content than you're used to.}

\subsection{Movie Clustering} \label{section 4.2}
MovieLens currently offers users four personalized recommendation algorithms to generate candidate movies: 1) "Peasant", a popularity-based non-personalized recommender that predicts rating from the average value of a user-item matrix, 2) "Warrior", an item-item Collaborative Filtering (CF) recommender that calculates movie predicted scores with cosine similarity among neighbor items \citep{sarwar2001item}, 3) "Wizard", which utilizes Funk Singular Value Decomposition (SVD) with 50 features and 125 training epochs per feature for matrix factorization \citep{funk2006netflix}, and 4) "Bard", an item-item CF based on user allocated points to different groups of movies \citep{chang2015using}. More detailed explanation of each base algorithms can be found in the paper by \citeauthor{ekstrand2015letting}. For this study, we introduced a hybrid approach by applying re-ranking logic on the top-generated recommendation candidates based on movie diversity clustering without changing any of the base Collaborative Filtering (CF) algorithms. The re-ranking procedure is detailed below.

\subsubsection {Re-ranking with K-Means} \label{re-ranking} There existed many diversity-based re-ranking algorithms, such as the usage of Maximal Marginal Relevance (MMR) \citep{carbonell1998use}, replacing top-N items with random infrequent candidates \citep{lathia2010temporal}, or employing Determinantal Point Process (DPP) \citep{wang2020personalized}. Though many granular options were available, our goal in this study was to 1) keep a consistent notion of diversity across the relatively inconsistent available items in the movie space and help users achieve different exploration goals, and 2) make the change of diversity level perceivable enough by end users so that their consequent behaviors can be observed. Rather than designing complicated logistics to find optimal distances between items and reach broadest coverage, we also needed to make sure the re-ranking of movies capture what users really want. Therefore, we were inspired by \citeauthor{chang2015using}'s and \citeauthor{kotkov2020clusterexplorer}'s clustering method and chose K-Means clustering \citep{hartigan1979algorithm} to fulfill the needs of selecting most diverse movies from a personalized recommendation set. We represented each movie with a \emph{Tag Genome} feature \citep{vig2012tag}, which is a vector with normalized relevance scores indicating different characteristics users used to describe the movie. The vector shape was [1, 1128]. For further details on the implementation logic, refer to the original paper by \citeauthor{vig2012tag}. All movies in the MovieLens database were clustered using the K-Means clustering algorithm \citep{hartigan1979algorithm}, with \emph{k} = 24. We experimented with different values of \emph{k} ranging between \numrange{12}{48} on 5 pilot users and found that 24 received the most preferences (with verbal input from pilot users to the question "Which recommendation list do you like most?"). To validate the cluster number choice, three researchers conducted offline analysis by comparing the top 10 tags (by relevance scores) of each cluster center. The analysis showed that increasing \emph{k} increased the redundancy of clusters, and \emph{k} = 24 gave a meaningful representation with little overlap.

After clustering the movies, we filtered candidate movies from a certain number of clusters corresponding to the user-selected diversity level. For the BRC group, the level was always fixed at 5, indicating the most diverse candidates were chosen from all 24 available clusters. Therefore, movies from all 24 clusters would be chosen to compose the recommendation list. For the BRC+DS group, we started with 5 most similar clusters at level 1 and increased by approximately 5 for every higher level until we reached level 5 with movies from 24 clusters. We then re-ranked the first 3-page recommendations of the top 72\footnote{We selected 72 to ensure real-time calculation feasibility and avoid affecting user experience with potential delay in recommendation serving time.} movies from the pool of personalized candidates generated by the base CF algorithm by greedily selecting movies that belonged to a chosen subset of clusters based on the user-defined diversity level. For example, for a BRC+DS user who selected level 3 as their desired recommendation diversity, they would see movies from 15 clusters on the first 3 pages of their recommendation. If they dragged the slider to level 5 and refreshed the page, they would see movies from all 24 clusters, with exactly one movie from each cluster on each of the first 3 pages. The greedy logic is presented in Algorithm \ref{greedy_algo}.

\begin{algorithm}
\caption{Greedy Cluster-based Re-ranking of Top Movies}
\begin{algorithmic}

\REQUIRE total cluster number = 24, user defined diversity level = $L$, $L \in \{1, 2, 3, 4, 5\}$ 
\ENSURE {the initial cluster $I$ is the one with highest rating counts from all users}
\STATE {subset\_cluster\_list = \{$I$\}}
\WHILE{size(subset\_cluster\_list) $ < \min(24, 5 * L)$}
    \STATE{Find a new cluster that can minimize the diversity of subset\_cluster\_list between all cluster centers based on the Diversity equation (\ref{eq:ils_formula})}
    \STATE{Append the new cluster to subset\_cluster\_list}
\ENDWHILE

\REQUIRE {candidate\_list = a list of movies sorted based on pre-calculated personalization score with base CF algorithms and each belonged to exactly one cluster}
\REQUIRE {max\_movies\_per\_cluster = ceiling(24 / (5 * L))}
\FOR{page in range \numrange{1}{3}}
    \STATE {curr\_page\_reranked\_list = \{\}}
    \WHILE{$size(curr\_page\_reranked\_list) < 24$}
    
        \STATE{Iterate each movie in candidate\_list, only append the movie to curr\_page\_reranked\_list if it belongs to a cluster $A$ in subset\_cluster\_list and size(movies added to curr\_page\_reranked\_list that also belongs to $A$) $<$ max\_movies\_per\_cluster}

    \ENDWHILE
\ENDFOR
\end{algorithmic}
\label{greedy_algo}
\end{algorithm}

\subsection{Measurement}
Inspired by \citeauthor{zhao2017toward} and \citeauthor{knijnenburg2012explaining}'s work for measuring user interactions with the system (INT), user-perceived subjective system aspects (SSA), and user experience (EXP) with the recommender~\citep{zhao2017toward,knijnenburg2012explaining}, we designed our log analysis metrics into the three categories we aimed to answer our \textbf{\emph{RQ3}} with: \textbf{Exploration}, \textbf{Engagement}, and \textbf{Satisfaction}. We tried to capture the metrics by both database logging as shown in Table \ref{log_analysis_metrics} and the post-survey questions shown in Table \ref{post_survey_metrics_table}. 

The post-survey questions were displayed on a 5-point Likert scale, except for the open-ended qualitative question. The first five questions of the survey corresponded to SSA metrics, and we used \emph{Ease of Use} and \emph{Satisfaction} to measure EXP. We classified each metric in Table \ref{post_survey_metrics_table} into one of the three categories mentioned above, based on the nature of the questions. The open-ended question aimed to gather users' overall experiences and their views on the usefulness of the new interface. We included users' feedback as quotes to triangulate the quantitative results and understand the reasons behind the quantitative data in later discussion. Finally, we designed two slider-interface-specific questions to assess how different users reacted to that interface.

\begin{table}
\centering
\resizebox{\linewidth}{!}{%
\begin{tabular}{lcc} 
\hline
\textbf{Category} & \textbf{Metric}    & \textbf{Description}                                                             \\ 
\hline
                  & ratingDiversity    & diversity measure~ on rated movies, computed with equation (\ref{eq:ils_formula})                    \\
Exploration       & sliderInteractions & slider Interaction count for groups under slider interface                       \\
                  & pageViewFreq       & number of unique movies user click to view from recommendation page per session  \\ 
\hline
                  & loginFrequency     & times users come to use the recommender per month                                \\
Engagement        & totalLength        & sum of all of a user's session length in minute                                  \\
                  & numRatings         & number of movies user rates                                                      \\ 
\hline
Satisfaction      & wishlistFreq       & the mean number of movies added to wishlist per login session~                   \\
                  & avgRating          & the mean ratings on all movies a user rates                                      \\
\hline
\end{tabular}
}
\caption{Log analysis metrics to quantify user interaction (INT) with the recommender}
\label{log_analysis_metrics}
\end{table}

\begin{table}
\centering
\resizebox{\linewidth}{!}{%
\begin{tabular}{cl} 
\toprule
\textbf{\textit{Metric}}                        & \textbf{Survey Question}                                                                                                                                                                                                                            \\ 
\hline
\textit{Accuracy (Satisfaction)}                & My recommendations in the new system match my taste in movies.                                                                                                                                                                                      \\ 
\hline
\textit{Diversity (Exploration)}                & \begin{tabular}[c]{@{}l@{}}Compared to previous MovieLens system, the new interface showed me:\\a narrower/broader set of movies.\end{tabular}                                                                                                      \\ 
\hline
\textit{Novelty (Exploration)}                  & \begin{tabular}[c]{@{}l@{}}Compared to previous MovieLens system, the new interface provided me:\\fewer movies I’ve not heard about / more movies I’ve not heard about.\end{tabular}                                                                \\ 
\hline
\textit{Level of Effort (Exploration)}          & \begin{tabular}[c]{@{}l@{}}Compared to previous MovieLens system, which interface helps you \\best discover movies you’d like to watch?\end{tabular}                                                                                                \\ 
\hline
\textit{Trustworthiness (Satisfaction)}         & \begin{tabular}[c]{@{}l@{}}Compared to previous MovieLens system, the new interface is: \\less trustworthy / more trustworthy\end{tabular}                                                                                                          \\ 
\hline
\textit{Ease of Use (Satisfaction)}             & The new interface is easy to follow and use                                                                                                                                                                                                         \\ 
\hline
\textit{Satisfaction (Satisfaction)}            & How satisfied are you with the changes?                                                                                                                                                                                                             \\ 
\hline
\textit{Usage Frequency (Engagement)}           & \begin{tabular}[c]{@{}l@{}}If you have both the old and new interfaces available,\\which would you use more?\end{tabular}                                                                                                                           \\ 
\hline
Qualitative                                     & \begin{tabular}[c]{@{}l@{}}Could you write a 2-3 sentence blurb to advertise this new versionof MovieLens \\to other users? How would you describe the new advantages of this new interface?\end{tabular}                                           \\ 
\hline\hline
\textit{Slider Group Specific Q1 (Exploration)} & What was your preferred slider level?                                                                                                                                                                                                               \\ 
\hline
\textit{Slider Group Specific Q2 (Exploration)} & \begin{tabular}[c]{@{}l@{}}Please tell us why you prefer this position? (choose all that apply) \\a. It had better variety \\b. It matched my taste better \\c. I did not see much difference between positions \\d. Others (specify)\end{tabular}  \\
\bottomrule
\end{tabular}
}
\caption{SSA and EXP metrics with corresponding survey questions}
\label{post_survey_metrics_table}
\end{table}

\section{Log Analysis}
To ensure the randomization of user selection in each treatment group (BRC, BRC+DS, Control), we conducted a pre-experiment one-way ANOVA statistical test~\citep{lazar2017research} on all INT metrics among the three interface treatments within each diversity group. This was done to obtain user interaction data exactly one month before the experiment, and we found no significant differences. We then applied the same ANOVA test to during-experiment and post-experiment data to compare the instant and longitudinal effectiveness of different interfaces on each diversity group.
For the post-experiment data, we collected user interaction logs for an additional three months after the experiment was terminated (from 12/17/2022 to 3/17/2023). To conduct post-hoc pairwise comparisons, we used a paired t-test with Cohen's $d$ \citep{cohen1988edition} as the effect size to measure the effect size between conditions within each diversity group. While we attempted to apply logistic normalization to different metrics, due to the different value scales of our data and the few significant differences observed, we decided to conduct between-subject and between-group analyses using the original interaction data. The basic mean values of different metrics within each factorial group is displayed in Table \ref{table4}.

\begin{table}
\centering
\resizebox{\linewidth}{!}{%
\begin{tabular}{cccccccccc} 
\hline
\multicolumn{2}{c}{\multirow{2}{*}{Group \textbackslash{} Metrics}} & \multicolumn{3}{c}{Exploration}                                                                                                                                                  & \multicolumn{3}{c}{Engagement}                                                                                                                                     & \multicolumn{2}{c}{Satisfaction}                                                                              \\
\multicolumn{2}{c}{}                                                & \begin{tabular}[c]{@{}c@{}}rating\\Diversity\end{tabular} & \begin{tabular}[c]{@{}c@{}}slider\\Interaction\end{tabular} & \begin{tabular}[c]{@{}c@{}}page\\ViewFreq\end{tabular} & \begin{tabular}[c]{@{}c@{}}login\\Freq\end{tabular} & \begin{tabular}[c]{@{}c@{}}total\\Length\end{tabular} & \begin{tabular}[c]{@{}c@{}}num\\Ratings\end{tabular} & \begin{tabular}[c]{@{}c@{}}wishlist\\Freq\end{tabular} & \begin{tabular}[c]{@{}c@{}}avg\\Rating\end{tabular}  \\ 
\hline
\multirow{2}{*}{D-BRC}      & During                                & 0.344                                                     & -                                                           & 2.505                                                  & 11.230                                              & 81.958                                                & 13.427                                               & 0.786                                                  & 2.864                                                \\
                            & Post                                  & 0.382                                                     & -                                                           & 2.466                                                  & 7.883                                               & 146.153                                               & 18.653                                               & 0.521                                                  & 2.706                                                \\
\multirow{2}{*}{D-BRC+DS}   & During                                & 0.364                                                     & 0.928                                                       & 2.337                                                  & 11.589                                              & 124.694                                               & 12.655                                               & 0.687                                                  & 2.824                                                \\
                            & Post                                  & 0.380                                                     & -                                                           & 2.212                                                  & 8.886                                               & 242.678                                               & 25.493                                               & 0.807                                                  & 2.766                                                \\
\multirow{2}{*}{D-Control}  & During                                & 0.346                                                     & -                                                           & 2.081                                                  & 10.601                                              & 86.032                                                & 9.235                                                & 1.034                                                  & 2.886                                                \\
                            & Post                                  & 0.388                                                     & -                                                           & 2.119                                                  & 7.943                                               & 172.559                                               & 22.940                                               & 0.877                                                  & 2.939                                                \\ 
\hline
\multirow{2}{*}{ND-BRC}     & During                                & 0.380                                                     & -                                                           & 1.956                                                  & 7.024                                               & 34.221                                                & 6.021                                                & 0.676                                                  & 2.931                                                \\
                            & Post                                  & 0.444                                                     & -                                                           & 2.070                                                  & 5.684                                               & 106.484                                               & 14.605                                               & 0.878                                                  & 2.847                                                \\
\multirow{2}{*}{ND-BRC+DS}  & During                                & 0.344                                                     & 0.668                                                       & 1.678                                                  & 8.348                                               & 65.016                                                & 6.699                                                & 0.543                                                  & 2.836                                                \\
                            & Post                                  & 0.426                                                     & -                                                           & 2.369                                                  & 6.171                                               & 132.718                                               & 13.774                                               & 0.675                                                  & 2.856                                                \\
\multirow{2}{*}{ND-Control} & During                                & 0.365                                                     & -                                                           & 1.997                                                  & 5.723                                               & 33.375                                                & 5.915                                                & 0.562                                                  & 2.767                                                \\
                            & Post                                  & 0.449                                                     & -                                                           & 1.730                                                  & 4.282                                               & 68.802                                                & 11.655                                               & 0.531                                                  & 2.847                                                \\
\hline
\end{tabular}
}
\caption{During and post experiment mean values of each INT metric on different user groups}
\label{table4}
\end{table}

\subsection{User Exploration Behavior Changes}

\begin{figure}[!htp]
    \centering
    \includegraphics[width=0.49\textwidth]{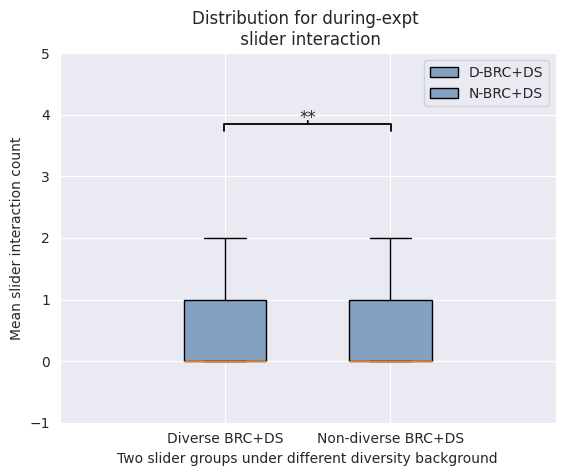}
    \includegraphics[width=0.49\textwidth]{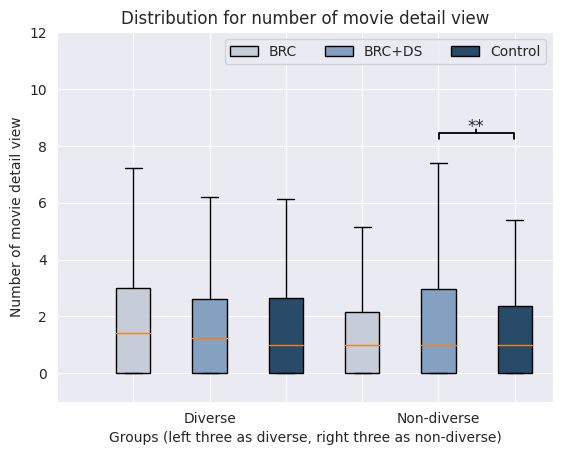}
    \caption{User exploration behavior log analysis. Asterisk (*) indicates a statistically significant difference between conditions:  $p<$ .1 (*); $p<$ .05 (**);  $p<$ .01 (***).}
    \label{log_analysis_exploration}
    \end{figure}

In terms of exploration, our analysis in Fig. \ref{log_analysis_exploration} revealed a significant difference between diverse and non-diverse BRC+DS user groups in the number of slider interactions during the experiment ($p_{(D-BRC+DS,ND-BRC+DS)} = 0.039$, $t_{(D-BRC+DS, ND-BRC+DS)} = 2.073$, $DOF = 583.495$, $cohen\_d_{(D-BRC+DS,ND-BRC+DS)} = 0.167$), as determined by a paired-t test. On average, diverse users had \textbf{\emph{0.928}} slider interactions, while non-diverse users had \textbf{\emph{0.668}} interactions. We also observed a similar pattern in the post-experiment movie detail page view count. Specifically, within the non-diverse user group, the BRC+DS interface motivated users to click and view more unique movie pages, even after the experiment was over and the new BRC+DS interface was removed ($p_{(ND-BRC+DS,ND-CT)} = 0.047$, $t_{(ND-BRC+DS,ND-CT)} = 1.993$, $DOF = 472.29$, $cohen\_d_{(ND-BRC+DS,ND-CT)} = 0.158$). Thus, the non-diverse user group's exploration behavior was positively impacted in a longitudinal way by the BRC+DS interface.

\subsection{User Engagement Pattern Changes}

\begin{figure}[!htp]
    \centering
    \includegraphics[width=0.49\textwidth]{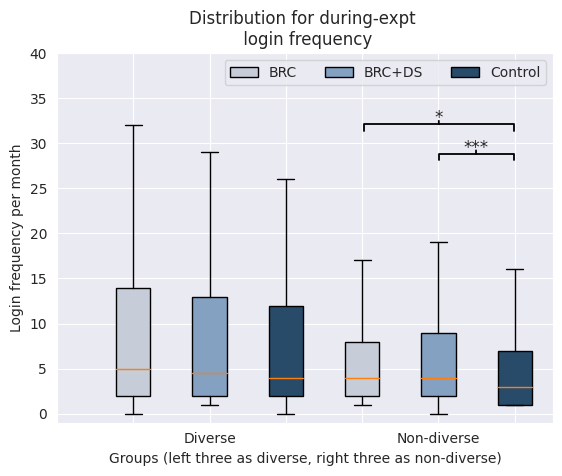}
    \includegraphics[width=0.49\textwidth]{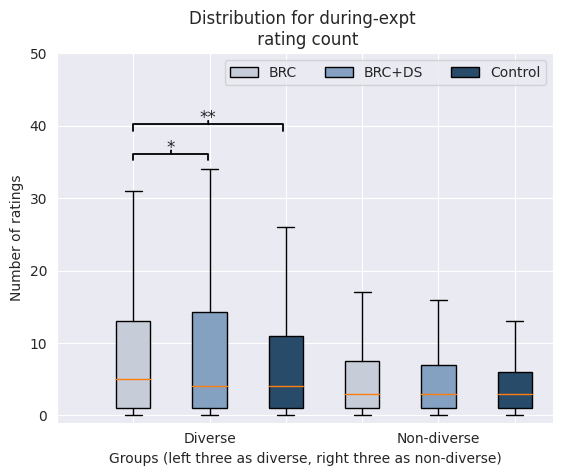}
    \caption{User engagement pattern changes during experiment. Asterisk (*) indicates a statistically significant difference between conditions:  $p<$ .1 (*); $p<$ .05 (**);  $p<$ .01 (***).}
    \label{log_analysis_engagement_during_expt}
    \end{figure}

Our 4 user engagement metrics all more or less revealed different engagement patterns of users under different exploration interfaces. When we look at during-experiment between-group changes in Fig. \ref{log_analysis_engagement_during_expt}, we see that login frequency of non-diverse users were positively impacted during experiment period, and the BRC+DS interface brought most significant changes on them ($p_{(ND-BRC+DS,ND-CT)} = 0.003$, $t_{(ND-BRC+DS,ND-CT)} = 3.024$, $DOF = 525.484$, $cohen\_d_{(ND-BRC+DS,ND-CT)} = 0.239$), followed by the BRC carousel ($p_{(ND-BRC,ND-CT)} = 0.056$, $t_{(ND-BRC,ND-CT)} = 1.915$, $DOF = 657.566$, $cohen\_d_{(ND-BRC+DS,ND-CT)} = 0.148$), both comparing to the control group. But in terms of new movie rating contribution, diverse user group was more heavily impacted compared to the non-diverse group. We witness a significant difference between the diverse BRC and control group ($p_{(D-BRC,D-CT)} = 0.02$, $t_{(D-BRC,D-CT)} = 2.333$, $DOF = 503.363$, $cohen\_d_{(D-BRC,D-CT)} = 0.19$), and also the fact that BRC interface attracts more users to rate than the BRC+DS interface ($p_{(D-BRC,D-BRC+DS)} = 0.06$, $t_{(D-BRC,D-BRC+DS)} = 1.889$, $DOF = 365.607$, $cohen\_d_{(D-BRC,D-BRC+DS)} = 0.157$). This finding is quite interesting since intuitively we would expect more engagement and potentially more rating contribution from users using the interactive BRC+DS interface, but the experiment findings are not aligned with that speculation.

\begin{figure}[!htp]
    \centering
    \includegraphics[width=0.49\textwidth]{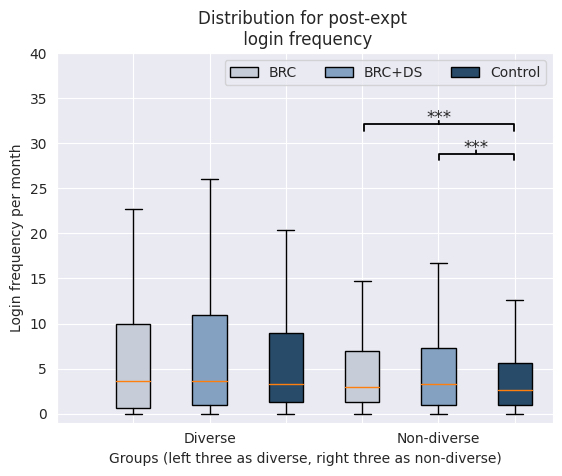}
    \includegraphics[width=0.49\textwidth]{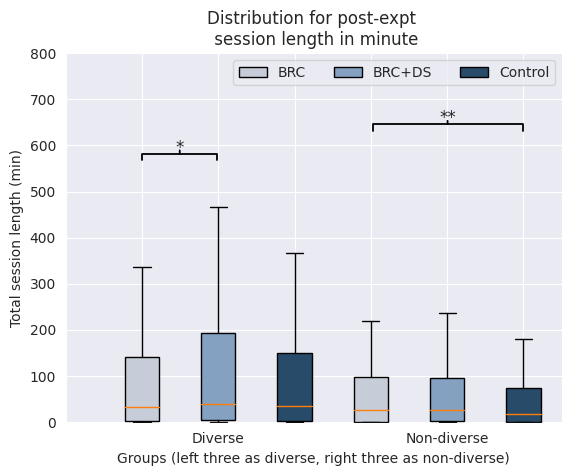}
    \caption{User engagement pattern changes post experiment. Asterisk (*) indicates a statistically significant difference between conditions:  $p<$ .1 (*); $p<$ .05 (**);  $p<$ .01 (***).}
    \label{log_analysis_engagement_post_expt}
    \end{figure}

In post-experiment data shown in Fig. \ref{log_analysis_engagement_post_expt}, we still found statistical significant difference within the non-diverse groups, following a similar pattern as we observed during experiment period that both BRC+DS ($p_{(ND-BRC+DS,ND-CT)} = 0.01$, $t_{(ND-BRC+DS,ND-CT)} = 2.579$, $DOF = 455.518$, $cohen\_d_{(ND-BRC+DS,ND-CT)} = 0.205$) and BRC ($p_{(ND-BRC,ND-CT)} = 0.009$, $t_{(ND-BRC,ND-CT)} = 2.622$, $DOF = 613.107$, $cohen\_d_{(ND-BRC+DS,ND-CT)} = 0.202$) interfaces generated positive impact on attracting users to login back to use the recommender. Additionally, in this period, we see some session length differences in both diverse and non-diverse user groups. For the BRC and BRC+DS diverse users, the interactive slider interfaces kept users to stay longer than just a broad recommendation carousel ($p_{(D-BRC,D-BRC+DS)} = 0.052$, $t_{(D-BRC,D-BRC+DS)} = -1.949$, $DOF = 409.588$, $cohen\_d_{(D-BRC,D-BRC+DS)} = -0.157$). While on the non-diverse group, BRC was bringing positive impact on user session length extension ($p_{(ND-BRC,ND-CT)} = 0.029$, $t_{(ND-BRC,ND-CT)} = 2.194$, $DOF = 566.62$, $cohen\_d_{(ND-BRC,ND-CT)} = 0.169$).

\subsection{User Satisfaction Metric Changes}

\begin{figure}[!htp]
    \centering
    \includegraphics[width=0.6\textwidth]{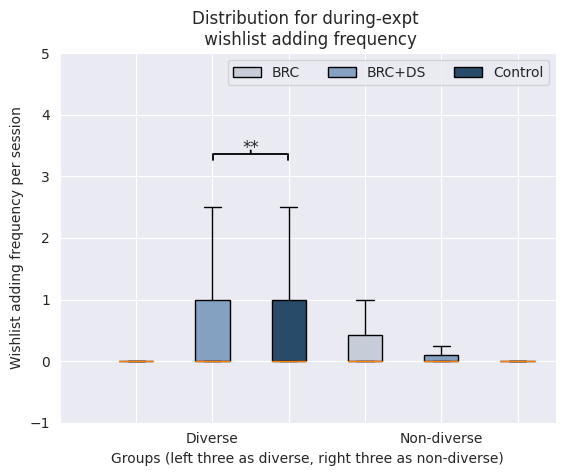}
    \caption{User satisfaction metric changes during experiment. Asterisk (*) indicates a statistically significant difference between conditions:  $p<$ .1 (*); $p<$ .05 (**);  $p<$ .01 (***).}
    \label{log_analysis_satisfaction_during_expt}
    \end{figure}

One direct outcome of users' exploration experience is their satisfaction level. Satisfaction metrics provided an angle for us to understand whether our new interface helped users find movies they would like to watch (therefore adding them to their wishlist), and if the recommendation quality was good in users' subjective opinions (reflected by the average ratings they gave to movies). During the experiment period (see Fig. \ref{log_analysis_satisfaction_during_expt}), we observed that the diverse BRC+DS users' frequency of adding movies to their wishlist actually decreased compared to the control group ($p_{(D-BRC+DS,D-CT)} = 0.047$, $t_{(D-BRC+DS,D-CT)} = -1.996$, $DOF = 483.236$, $cohen_\_d_{(D-BRC+DS,D-CT)} = -0.168$). 
This finding indicates that the BRC+DS interface had a negative impact on helping diverse users find movies they wanted to watch later. 

\begin{figure}[!htp]
    \centering
    \includegraphics[width=0.49\textwidth]{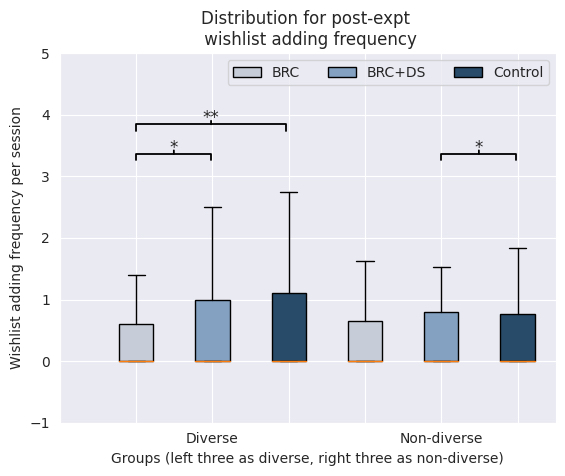}
    \includegraphics[width=0.49\textwidth]{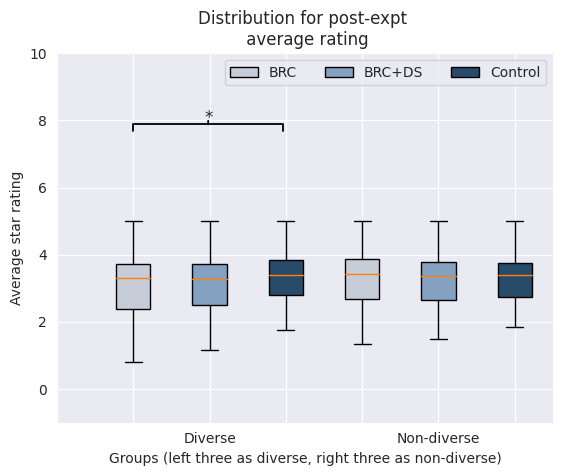}
    \caption{User satisfaction metric changes post experiment. Asterisk (*) indicates a statistically significant difference between conditions:  $p<$ .1 (*); $p<$ .05 (**);  $p<$ .01 (***).}
    \label{log_analysis_satisfaction_post_expt}
    \end{figure}

During the post-experiment period, we observed a similar detrimental impact on wishlist adding and average rating for the diverse group of users, as shown in Fig. \ref{log_analysis_satisfaction_post_expt}. The BRC interface had the most negative influence compared to both the control group ($p_{(D-BRC,D-CT)} = 0.018$, $t_{(D-BRC,D-CT)} = -2.373$, $DOF = 511.32$, $cohen\_d_{(D-BRC,D-CT)} = -0.2$) and the BRC+DS group ($p_{(D-BRC,D-BRC+DS)} = 0.087$, $t_{(D-BRC,D-BRC+DS)} = -1.717$, $DOF = 502.436$, $cohen\_d_{(D-BRC,D-BRC+DS)} = -0.139$). In addition, the average ratings from BRC diverse users were lower compared to the control group ($p_{(D-BRC,D-CT)} = 0.057$, $t_{(D-BRC,D-CT)} = -1.906$, $DOF = 564.209$, $cohen\_d_{(D-BRC,D-CT)} = -0.16$). However, the non-diverse group of users seemed to find the BRC interface useful in helping them discover more movies to watch later, compared to the control setting ($p_{(ND-BRC,ND-CT)} = 0.097$, $t_{(ND-BRC,ND-CT)} = 1.662$, $DOF = 403.55$, $cohen\_d_{(ND-BRC,ND-CT)} = 0.127$).

\section{Post Survey Results}

Table \ref{post_survey_mean_vals} presents the mean values of the original quantitative post-survey input. To categorize the results more systematically, we first converted the answers of the 5-Likert scale into three ordinal categories: negative (including both 1 and 2 points on the scale), neutral (corresponding to 3 points on the scale), and positive (corresponding to 4 and 5 points on the scale). In order to investigate whether any survey responses can better predict user-perceived satisfaction with the system, we ran the ordinal logistic regression (OLR) analysis \citep{kleinbaum2010ordinal} with logistic method on four different treatment groups (D-BRC, D-BRC+DS, ND-BRC, ND-BRC+DS) to check the potential contribution from each quantitative user subjective metric to the final satisfaction. We also transformed the Interface (Control $->$ 0, BRC $->$ 1, BRC+DS $->$ 2) and user consumption habit (D $->$ 1, ND $->$ 2) and included both in the regression analysis.

\begin{table}
\centering
\resizebox{\linewidth}{!}{%
\begin{tabular}{cccccccc} 
\hline
\multicolumn{2}{c}{\multirow{2}{*}{Metrics \textbackslash{} Group}} & D-BRC  & D-BRC+DS & D-CT   & ND-BRC & ND-BRC+DS & ND-CT   \\
\multicolumn{2}{c}{}                                                & (n=38) & (n=41)   & (n=35) & (n=39) & (n=34)    & (n=32)  \\ 
\hline
             & Diversity                                            & 3.632  & 3.537    & 3.471  & 3.359  & 3.529     & 3.067   \\
Exploration  & Novelty                                              & 3.289  & 3.366    & 3.588  & 3.103  & 3.353     & 3.233   \\
             & Level of Effort                                      & 3.368  & 3.341    & 3.324  & 3.308  & 3.382     & 3.167   \\
             & Slider Group Q1                                      & -      & 3.22     & -      & -      & 3.182     & -       \\ 
\hline
             & Trustworthiness                                      & 3.237  & 3.122    & 3.206  & 3.179  & 3.176     & 3.067   \\
Satisfaction & Ease of Use                                          & 3.868  & 3.902    & 4.086  & 3.872  & 3.853     & 3.7     \\
             & Satisfaction                                         & 3.658  & 3.61     & 3.5    & 3.436  & 3.324     & 3.2     \\
             & Accuracy                                             & 3.711  & 3.781    & 3.657  & 3.718  & 3.706     & 3.25    \\ 
\hline
Engagement   & Usage Frequency                                      & 3.5    & 3.25     & 3.588  & 3.41   & 3.636     & 3.345   \\
\hline
\end{tabular}
}
\caption{Mean values of quantitative survey questions}
\label{post_survey_mean_vals}
\end{table}

As shown in Table \ref{post_survey_coef_table}, \emph{Usage frequency} is positively correlated with \emph{Satisfaction} with significance. Though \emph{Accuracy} and \emph{Diversity} responses have higher coefficients, they do not show significance. On the contrary, \emph{Interface} and \emph{Consumption habit} play negative roles on significantly impacting user satisfaction. As we denoted BRC with unit 1 and BRC+DS as unit 2, that indicates users actually enjoyed the BRC exploration mode more. Also, we denoted diverse users as 1 and non-diverse users as 2, which means diverse users are more satisfied with our interface changes than non-diverse users. The results from Odds ratio that is generated by the coefficient and confidence intervals between 2.5\% and 97.5\% also support our observation. For those users who preferred to use original interface (as captured by the \emph{Usage frequency} question), the odds of being less satisfied was 0.898 times that of users who preferred to use the new exploration interface. Also, the odds ratio suggested that BRC+DS interface was 0.709 times less likely to make users satisfied compared to the BRC interface. 

\begin{table}
\centering
\resizebox{\linewidth}{!}{%
\begin{tabular}{llllllll} 
\hline
                   & Coef            & Std. Error & t value & p value             & Odds ratio         & 2.5 \%          & 97.5 \%          \\
Interface          & \textbf{-0.709} & 0.357      & -1.985  & \textbf{4.72E-02**} & \textbf{-7.09E-01} & \textbf{-1.421} & \textbf{-0.016}  \\
Consumption\_habit & \textbf{-1.022} & 0.365      & -2.800  & \textbf{5.10E-03**} & \textbf{-1.02E+00} & \textbf{-1.755} & \textbf{-0.319}  \\
Accuracy           & \textbf{0.261}  & 0.338      & 0.774   & 4.39E-01            & 2.61E-01           & -0.397          & 0.934            \\
Diversity          & \textbf{0.284}  & 0.296      & 0.959   & 3.37E-01            & 2.84E-01           & -0.296          & 0.872            \\
Novelty            & 0.144           & 0.256      & 0.565   & 5.72E-01            & 1.45E-01           & -0.359          & 0.650            \\
Level\_of\_effort  & -0.002          & 0.401      & -0.006  & 9.95E-01            & -2.28E-03          & -0.796          & 0.789            \\
Trustworthiness    & 0.061           & 0.376      & 0.162   & 8.71E-01            & 6.10E-02           & -0.679          & 0.803            \\
Ease\_of\_use      & -0.076          & 0.336      & -0.227  & 8.20E-01            & -7.63E-02          & -0.734          & 0.591            \\
Usage\_frequency   & \textbf{0.898}  & 0.341      & 2.630   & \textbf{8.54E-03**} & \textbf{8.98E-01}  & \textbf{0.235}  & \textbf{1.582}   \\
\hline
\end{tabular}
}
\caption{OLR Coefficient and significance of each survey question to satisfaction. Significant p-values are denoted with **. }
\label{post_survey_coef_table}
\end{table}

\section{Discussion}

In this section, we integrate both user log interaction and survey input data to answer our final research question:

\textbf{\emph{RQ3}}: How can the new recommendation interfaces affect users' exploration, engagement, and satisfaction for those with different diversity consumption habits?

\subsection{Impact on Exploration}
When examining the user exploration patterns, it becomes apparent that non-diverse users benefited more from the new exploration interface, especially in terms of broadening their recommendation preferences. For instance, ND users were more likely to click on unique movies during the post-experiment session when using the BRC+DS interface (Fig. \ref{log_analysis_exploration}). One non-diverse user mentioned in their survey feedback, \emph{"[the recommender] provided me with a service to reliably find new movies which combined with the statistics provided about user's ratings motivated me to watch more movies and more importantly seek out movies from directories I haven't explored much yet"}. In other words, the exploration interfaces motivated users to step out of their comfort zone and mitigate the pigeonholing effect. Regarding diverse user groups, although there was no interaction data to provide insights, one user praised the BRC interface for its diversity advantage, stating that it \emph{"helps you break out of your filter bubble."} or for its recommendation novelty, which \emph{"helps [me] find more off the radar movies that I would not have easily found on my own."}

\subsection{Impact on Engagement}

The more active exploration of new interface features by diverse users led to more positive engagement behavior (Fig. \ref{log_analysis_exploration}). Specifically, the BRC interface attracted higher login frequency and more rating counts from diverse users during and post the experiment period, even outperforming the BRC+DS interface in terms of promoting more rating counts from users during the experiment period. The BRC+DS interface also encouraged more logins and longer session duration for diverse users. On the other hand, non-diverse users only extended their session length with the BRC interface. Overall, users who historically consumed and explored more actively were more likely to engage with the new interfaces and the entire recommender system. This could be explained by the fact that diverse users had more exposure and potentially more motivation to try out new features.

\subsection{Impact on Satisfaction}

Although the positive impact on diverse user engagement was observed, a negative impact on their wishlist addition and average rating was also noticed. On the other hand, for the non-diverse group, the BRC interface had a positive impact on wishlist addition post-experiment. More insights from the regression analysis in Table \ref{post_survey_coef_table} indicated that both exploration interface (captured by the \emph{Interface} coefficient) and users' own preference on old or new interface to use (captured by the \emph{Usage frequency coefficient}) served as a good indicator of their satisfaction. An interesting fact we observed is that, BRC as a less interactive interface actually gained more user satisfaction than BRC+DS (captured by the \emph{Interface} odds ratio), which implied that users preferred to get pre-generated diverse recommendations directly, instead of adjusting sliders to figure out what they like on their own. 

However, it is important to note that the general patterns from group data may not always provide a representative picture for each individual user. For instance, one D-BRC+DS user said, \emph{"The new interface definitely seems more honed in on my personal tastes. A solid improvement on my preferred way to find new movies to watch."}. Meanwhile, an ND-BRC user complained:  \emph{"...the broad recommendations didn't do it for me -- I suspect because the confidence was low: I saw that the movies recommended 'most' were still movies for which the prediction was less than 3. I'd rather skip those."}. Since MovieLens and many other similar recommender websites have the "predicted rating" feature, users might have built their confidence on it, which can easily impact their overall impression and satisfaction with the system. Although the \emph{Trustworthiness} question did not strong contribute to user satisfaction in our survey regression analysis, this potential relationship can be an interesting area for future research.

Integrated with the findings from the two subsections above, we observe a comprehensive pattern that connects all three perspectives that we care about: both diverse and non-diverse users could differentiate various levels of recommendation broadness and found exploration assistance useful for widening their consumption, while BRC actually performed better than BRC+DS in terms of user perceived satisfaction. Diverse users were more engaged, but the majority of their recommendation diversity seems to be less impacted by the new interfaces. Regarding satisfaction, recommendation accuracy and diversity of content were not the best predictor of their satisfaction. Instead, satisfaction is a multi-dimensional metric that can be related to user historical consumption habits, the level of control on interface exploration, and users' own preference for the recommendation interface. But even in the user group whose general satisfaction metric was undermined, we still see certain outlier users who expressed appreciation. However, we admit that using a single metric to divide different user groups is still limited and cannot adequately represent the various exploration goals or consumption patterns of users.

\section{Conclusion and Future Implications}

In this study, we analyzed the current exploration pain points of human-recommender interaction and designed a human-in-the-loop content generation interface that enabled a more interactive and user-centric recommender system. At a high level, users appreciate access to more diverse recommendations, and the new interactive interfaces allow users to explore more broadly, engage more actively, and receive both personalized and accurate recommendations. We also revealed the correlation between interface exploration control level, users' consumption habit, and users' subjective preference on interface to their satisfaction.

To provide better design implications for future recommender systems, we argue that users' exploration patterns cannot be generalized with a single diversity threshold. Similarly, the same standard should be applied to assess their satisfaction, which is a multi-dimensional objective. Therefore, when designing new recommendation interfaces, designers and practitioners should make an effort to understand user needs with a representative sample and frequently adjust the recommendation strategy to accommodate their various exploration goals. For this purpose, we suggest using conversational recommenders \citep{jannach2021survey}, multi-objective personalized feed \citep{kong2022multi}, or AI-supported tools such as multi-armed bandit algorithms \citep{silva2022multi} to fulfill personalized and timely recommendation needs. However, it is crucial to keep in mind that all changes and strategies should align with users' trust and usage habits with the system.

We acknowledge some limitations in our study: 1) Our focus was mainly on active and long-term users, and we did not include new users in our analysis; 2) We used a single diversity threshold to categorize user exploration patterns and analyze user exploration diversity, which may not be the optimal approach and could lead to limitations in finding between-subject significance, as demonstrated by previous research \citep{ferwerda2016influence,tang2022preference,karumur2018personality,jesse2022intra,ge2013bringing}; 3) Our implementation of the greedy re-ranking algorithm is straightforward and only mean to deliver the most direct and efficient diversity shift purpose for later user behavior evaluation. We have no valid reason to support that it is the best option, and future refinement studies can further optimize it or try more granular re-ranking mechanisms for diversification purpose. 4) We only investigated the recommendation system in the movie domain and did not explore the possibility of expanding it to other domains, such as music and news recommendations; 5) Due to technical limitations, we were not able to deploy additional interfaces that users expressed interest in during the formative interview; 6) Even if we tried to represent the real-world interaction scenario as much as possible, our experiment was still conducted under a controlled environment with pre-selected users for between-group comparison.  

Looking forward, we hope that this study can provide a solid foundation for designing more human-centric recommender systems that can recognize and accommodate different user exploration needs, beyond accuracy or even diversity. Our focus in this study is to reveal the impact of user exploration on engagement and satisfaction measurement, and there is still plenty of space for further analysis on other pairwise or integrated correlations between the list of metrics we proposed. Besides, future studies can implant this mixed-methodology design to evaluate other important factors such as novelty, accountability, or general usability evaluation of recommender systems. We believe that understanding those different user-centric aspects can shed more light on the success of recommenders and lead to interesting follow-up studies in the future.

\bibliographystyle{apacite}
\bibliography{ref}

\appendix
\section*{Appendices} \label{appendix}

\setcounter{table}{0}
\renewcommand{\thetable}{A\arabic{table}}

\begin{longtable}{>{\hspace{0pt}}m{0.035\linewidth}>{\hspace{0pt}}m{0.879\linewidth}>{\hspace{0pt}}m{0.025\linewidth}} 
\cline{1-2}
\textbf{\textit{Q1}} & How often do you look at the recommendations? How often do you look at specific movies that are recommended on MovieLens?                                                                                                                                                                                                                                                                                                                                                                                                                                                                                                                                                                                                                                                                                                                                                                                                                                                                                                                                                                                             & \multirow{14}{0.025\linewidth}{\hspace{0pt}}  \\* 
\cline{1-2}
\textbf{\textit{Q2}} & When you come to MovieLens, which of the following goals and activities do you use the site for?~ Please mark all of the ones you commonly do.                                                                                                                                                                                                                                                                                                                                                                                                                                                                                                                                                                                                                                                                                                                                                                                                                                                                                                                                                                        &                                               \\*
                     & \textbf{\textit{(a) Finding Movies}}                                                                                                                                                                                                                                                                                                                                                                                                                                                                                                                                                                                                                                                                                                                                                                                                                                                                                                                                                                                                                                                                                  &                                               \\*
                     & ~ ~ [1]~\textcolor[rgb]{0.125,0.125,0.133}{Get suggestions for movies you already know but haven't thought of.~ ~ [2]~Find movies you're pretty sure you will like.}\par{}\textcolor[rgb]{0.125,0.125,0.133}{}~ ~ [3]~\textcolor[rgb]{0.125,0.125,0.133}{Find movies that are similar to other movies you've seen.}                                                                                                                                                                                                                                                                                                                                                                                                                                                                                                                                                                                                                                                                                                                                                                                                   &                                               \\*
                     & \textbf{\textit{(b)~\textcolor[rgb]{0.125,0.125,0.133}{\textbf{\textit{Finding different movies}}}}}                                                                                                                                                                                                                                                                                                                                                                                                                                                                                                                                                                                                                                                                                                                                                                                                                                                                                                                                                                                                                  &                                               \\*
                     & ~ ~ [1]~\textcolor[rgb]{0.125,0.125,0.133}{Find movies you have not heard of.}\par{}\textcolor[rgb]{0.125,0.125,0.133}{}~ ~ [2]~\textcolor[rgb]{0.125,0.125,0.133}{Find movies you might not usually like but you will enjoy trying.~ ~ [3]~Find movies that are different from what you usually watch.}                                                                                                                                                                                                                                                                                                                                                                                                                                                                                                                                                                                                                                                                                                                                                                                                              &                                               \\*
                     & \textcolor[rgb]{0.125,0.125,0.133}{\textbf{\textit{(c)~\textbf{\textit{Finding movies that fit}}}}}                                                                                                                                                                                                                                                                                                                                                                                                                                                                                                                                                                                                                                                                                                                                                                                                                                                                                                                                                                                                                   &                                               \\*
                     & \textcolor[rgb]{0.125,0.125,0.133}{~ ~ [1]~Find movies that match certain needs of yours (e.g., length, on TV, on streaming, showing in a theater near me).}\par{}\textcolor[rgb]{0.125,0.125,0.133}{~ ~ [2]~Find movies that match certain of your preference attributes (e.g., genre, director, actors, topic).}                                                                                                                                                                                                                                                                                                                                                                                                                                                                                                                                                                                                                                                                                                                                                                                                    &                                               \\* 
\cline{1-2}
\textbf{\textit{Q3}} & \textcolor[rgb]{0.125,0.125,0.133}{When browsing recommendations from the MovieLens main page, which carousel(s) do you think are most helpful and do you most commonly use?}\par{}\textcolor[rgb]{0.125,0.125,0.133}{~ ~ [a]~Top picks~ ~ [b]~Recent releases~ ~ [c]~Rate more~ ~ [d]~Favorite from the past year~ ~ [e]~New additions\textbf{}}                                                                                                                                                                                                                                                                                                                                                                                                                                                                                                                                                                                                                                                                                                                                                                     &                                               \\* 
\cline{1-2}
\textbf{\textit{Q4}} & \textcolor[rgb]{0.125,0.125,0.133}{Is there anything about MovieLens recommendations that don’t match your expectations? Do you have any pain points or frustrations with the site’s recommendations or the site as a whole?}                                                                                                                                                                                                                                                                                                                                                                                                                                                                                                                                                                                                                                                                                                                                                                                                                                                                                         &                                               \\* 
\cline{1-2}
\textbf{\textit{Q5}} & \textcolor[rgb]{0.125,0.125,0.133}{Sometimes recommender systems get “stuck” giving users too many recommendations from a narrow category such as their favorite genre, cast members, years, or tags. Have you experienced this phenomenon? Has it happened often? Did you feel it was good or bad for you? What type of narrow category did you find your recommendations stuck in?}\par{}\textcolor[rgb]{0.125,0.125,0.133}{~ ~ *~If they said no – follow up with “Do you think it would be useful to be able to get narrower recommendations to explore more choices that are similar to each other?”\textbf{}}                                                                                                                                                                                                                                                                                                                                                                                                                                                                                                   &                                               \\* 
\cline{1-2}
\textbf{\textit{Q6}} & Which MovieLens features do you find most useful?~ Are there features you don’t like?~ Are there any that you wish the site had?                                                                                                                                                                                                                                                                                                                                                                                                                                                                                                                                                                                                                                                                                                                                                                                                                                                                                                                                                                                      &                                               \\* 
\cline{1-2}
\textbf{\textit{Q7}} & MovieLens recommendations are based on your movie ratings, but otherwise, you have little control over the recommendations.~ Would you want to have more control – for example, choosing how broad or narrow the set of movies recommended would be?~ Or getting recommendations based on a set of movies you select, rather than based on your long-term rating profile?~ What types of controls would be interesting to you? Why? Why do you prefer some choices over others?                                                                                                                                                                                                                                                                                                                                                                                                                                                                                                                                                                                                                                       &                                               \\*
                     & ~ ~ [a]~Controls that provide you with information about the recommendation (e.g., how~broad or narrow they are) but not changing them.\par{}~ ~ [b]~A control to set how broad or narrow your recommendations are.\par{}~ ~ [c]~A slider where you could set: ~ Narrow~ \textless{}———————————\textgreater{} Broad, maybe along with a bunch of categories to toggle on.\par{}~ ~ [d]~Pop-up warnings when you’re being very narrow in what you’re looking at.\par{}~ ~ [e]~A way to base your recommendations on something other than your rating profile. For instance, just seeing the most popular movies, or seeing movies that match a list you could specify.\par{}~ ~ [f]~The options to specify filters as part of recommendation – for example, only recommend movies in theaters or movies rated PG-13 or lower.\par{}~ ~ [g]~Are there different carousels that would be useful for you?~ What would you want?~ How about a carousel focused on “top choices you might not have thought about”?\par{}~ ~ [h]~Other ideas you have about what would make MovieLens recommendations better or more useful? &                                               \\
\cline{1-2}
\caption{Formative interview scripts}
\end{longtable}

\section*{Author Biography}

\begin{description}
    \item[Ruixuan Sun] is a third-year Computer Science Ph.D. student at the GroupLens Lab of University of Minnesota, Twin Cities. Her research interests lie in the intersection of Human-Computer Interaction and Recommender Systems. She is passionate about explainable AI, social computing, and human-centric recommender system design and evaluations.
    
    \item[Avinash Akella] graduated from the University of Minnesota with a Master’s degree in Data Science in 2023. His research interests lie at the intersection of Human-Computer Interaction, Data Science, Computer Science, and Psychology. He is now a Data Scientist at IBM.

    \item[Ruoyan Kong] graduated from the University of Minnesota, Twin Cities with a Ph.D. in Computer Science in 2023. Her research interests include recommender systems, email personalization, and social computing. She is now a machine learning engineer at TikTok.

    \item[Moyan Zhou] is a third-year Ph.D. student in Computer Science at the University of Minnesota, Twin Cities in GroupLens Lab. Her research is focused on Human Computer Interaction and Peer Production and Online Collaboration Communities. 

    \item[Joseph A. Konstan] is Distinguished McKnight Professor of Computer Science and Engineering at the University of Minnesota, where he also serves as Associate Dean for Research in the College of Science and Engineering.  He has been working in the field of Recommender Systems since 1995, mostly with a focus on human-centered design and evaluation of recommender algorithms and interfaces.
    
\end{description}

\end{document}